%% file: mobaku2013.tex
\documentclass[a4paper,12pt]{article}

\usepackage[margin=2.8cm]{geometry} 
\usepackage{fleqn} 

\usepackage{amssymb}
\usepackage{amsmath}
\usepackage{amsfonts}
\usepackage{amsthm}

\usepackage{cite}              
\usepackage{url}
\usepackage{enumerate}

\usepackage[T1]{fontenc}
\usepackage[latin1]{inputenc} 

\usepackage{multicol}
\usepackage{mathrsfs}

\usepackage{import} 
\usepackage{graphics}
\usepackage{graphicx} 
\usepackage{overpic}
\usepackage{epsfig}
\usepackage{subfigure}
\usepackage{float}
\usepackage{epstopdf}
\graphicspath{/plots/}
\usepackage[usenames]{color}

\newcommand{\mbb}[1]{\mathbb{#1}}

\newcommand{\R}{\mbb{R}}

\newcommand{\taup}{\tau_p}

\newcommand{\mupj}{\mu_{p,1}}
\newcommand{\mupa}{\mu_{p,2}}

\newcommand{\murj}{\mu_{r,1}}

\newcommand{\bpj}{b_{p,1}}
\newcommand{\bpa}{b_{p,2}}
\newcommand{\bpp}{b_{p,3}}

\newcommand{\restr}[2]{\left.\kern-\nulldelimiterspace {#1}
\vphantom{\big|}\right|_{#2} }

\renewcommand{\epsilon}{\varepsilon}
\renewcommand{\theta}{\vartheta}

\newtheorem{theorem}{Theorem}[section]
\newtheorem{proposition}{Proposition}[section]
\newtheorem*{proposition*}{Proposition}

\newtheorem*{result*}{Result}

\theoremstyle{definition}

\newtheorem*{example*}{Example}
\newtheorem*{note*}{Remark}

\DeclareMathOperator{\tr}{trace}

\title{Predator-Prey Interactions, Age Structures and Delay Equations}
\author{Marcel Mohr\thanks{University of Heidelberg, Institute of Applied Mathematics, Im Neuenheimer Feld 295, D-69120 Heidelberg,
Germany, \textit{marcel.mohr@bioquant.uni-heidelberg.de}} \and Maria Vittoria
Barbarossa\thanks{Bolyai Institute,
University of Szeged, H-6720 Szeged, Aradi v\'{e}rtan\'{u}k tere 1, Hungary,
\textit{barbarossamv@gmail.com}} \and Christina Kuttler\thanks{Institute of
Mathematics,
Department for Mathematical
Modeling, Technische Universit\"at M\"unchen, Boltzmannstr. 3, D-85748 Garching
b. M\"unchen, Germany, 
\textit{kuttler@ma.tum.de}}}

\small{\date{Submitted on August 10th, 2013}}

\begin{document}

\maketitle

\begin{abstract}
A general framework for age-structured predator-prey systems is introduced. 
Individuals are distinguished into two classes, juveniles and adults, and
several possible interactions are considered. The initial system of partial
differential equations is reduced to a system of (neutral) delay differential
equations with one or two delays. Thanks to this approach,
physically correct models for predator-prey with delay are provided. Previous
models are 
considered and analysed in view of the above results. A Rosenzweig-MacArthur
model with delay is presented as an example.
\end{abstract}


\include{intro}
\include{model}
\include{analysis_modelchanged}

%

\end{document}

%% file: intro.tex
\section{Introduction}
\label{sec:intro}
One of the classical topics of population dynamics is the
description of predator-prey interactions. Historically, the first mathematical
approach for predator-prey dynamics was given by Lotka \cite{Lotka1925}
and Volterra \cite{Volterra1926} who proposed the system
\begin{equation}
\begin{aligned}
R'(t) &= b_rR(t) -\mu_rP(t) R(t), \\
P'(t) &= -\mu_pP(t) + b_p R(t) P(t). 
\end{aligned}
\label{sys:lotkavolterra}
\end{equation}
Here \(R(t)\) and \(P(t)\) denote the population size at time \(t\geq 0\) of
prey and predators, respectively. According to this model, in absence of
predators the prey population grows exponentially with rate $b_r>0$. In absence
of prey, the predator population decreases exponentially with rate $\mu_p>0$.
Prey encounter predators and are killed at rate $\mu_r>0$, consequently the
predator population increases with rate $b_p>0$. Solutions of
\eqref{sys:lotkavolterra} oscillate periodically about a nonnegative coexistence
point \cite{EdelsteinKeshet1988}.\\
\ \\
More realistic models than \eqref{sys:lotkavolterra} have been proposed. For
example, May \cite{May1973} suggested that the prey population grows
logistically, rather than exponentially,
\begin{equation}
\begin{aligned}
R'(t) &= b_rR(t)\left(1-\frac{R(t)}{K}\right) -\mu_rP(t) R(t), \\
P'(t) &= -\mu_pP(t) + b_p R(t)P(t). 
\end{aligned}
\label{sys:may1973}
\end{equation}
Stability analysis of
\eqref{sys:may1973} shows that the coexistence equilibrium, which depends also
on the carrying capacity $K$ of the prey population, is always a stable point
\cite{EdelsteinKeshet1988}. Including the assumption that the predation
underlies a saturation, one obtains a system
which is usually attributed to Rosenzweig and
MacArthur \cite{Rosenzweig1963}, namely
\begin{equation}
\begin{aligned}
R'(t) &= b_rR(t)\left(1-\frac{R(t)}{K}\right) -\frac{\mu_rP(t)R(t)}{1+R(t)}, \\
P'(t) &= -\mu_pP(t) + \frac{b_pP(t)R(t)}{1+R(t)}. 
\end{aligned}
\label{sys:rosenzweig}
\end{equation}
This system reproduces the so-called \textit{paradox of enrichment}: If the
carrying capacity $K$ of the prey is small, the coexistence point (when it
exists) is stable. For $K\to \infty$ a Hopf bifurcation occurs,
the coexistence point becomes unstable and a stable periodic orbit appears
\cite{EdelsteinKeshet1988}.\\
\indent In general a predator-prey system has the structure
\begin{equation}
\begin{aligned}
R'(t) &= \alpha(R(t))R(t) -g(R(t), P(t))P(t), \\
P'(t) &= -\gamma(P(t))P(t) +h(R(t), P(t))P(t), 
\end{aligned}
\label{sys:predpreygeneral}
\end{equation}
where $\alpha$ (respectively, $h$) is the growth rate
and $g$ (respectively, $\gamma$) is the death rate for the prey (the
predators). Several biological
experiments suggest that death of prey and birth of predators depend on the
total prey and/or
predator population size \cite{Abrams2000}. The function $g$ is usually
called \textit{the predation response function} and describes
how probable the prey dies due to predation. In general one assumes that $g:
[0,\infty) \times
[0,\infty) \to [0,\infty)$ is continuous and $g(0,0)=0$. The simplest $g(R, P)$
is a linear
function of $R$. For example in \eqref{sys:lotkavolterra} and
\eqref{sys:may1973}, we
find $g(R,P)=\mu_r R$, with $\mu_r>0$, the rate at which a prey
encounters a predator. A nonlinear choice for $g$ is also possible. For example,
a Holling type II
function, $g(R,P)=\frac{d R}{T+R}$, as in system
\eqref{sys:rosenzweig}, indicates that
predators can only eat up a limited number
of prey, whereas the function $g(R,P)=\frac{d R P}{T+RP}$ suggests that the
number of encounters
between predator and prey depends also on $P$. The function $h$ is also
called \textit{the fecundity response function} and describes how
the reproduction rate of predators depends on predation. It is common to let $h:
[0,\infty) \times [0,\infty) \to [0,\infty)$ be continuous, with $h(0,0)=0$.
However, the
biology suggests that there must be a certain relation between $h$ and $g$, and
in general
$h(R,P)=cg(R,P)$, with some  $c>0$, as in systems \eqref{sys:lotkavolterra},
\eqref{sys:may1973}, \eqref{sys:rosenzweig}.\\
\ \\
\noindent To model predator-prey interactions and include more details, one can
use partial
differential equations (PDEs) and consider age-structured populations
\cite{Cushing1982}. Alternative modelling approaches use systems of ordinary
differential equations (ODEs) or delay differential equations (DDEs) with
constant delays and suggest a sort of age structure, identifying several
maturity classes. The simplest models consider only two classes of
individuals, namely juvenile and adult ones \cite{Ross1972, Hastings1984,
Nunney1985a, Bartlett1957}.

\noindent In this paper we want to combine a PDE system with a DDE system and
show the connections between the two modelling approaches, as it was
done for single population dynamics, e.g., in \cite{Bocharov2000,Nisbet1983}.
The paper is organised as follows. In Section \ref{sec:model} we consider
age-structured prey and predator populations and define a threshold
age, \textit{age-at-maturity}, to
distinguish juvenile from adult individuals. A newborn individual enters the
juvenile class and if it survives up to the age-at-maturity it
enters the adult class. In our assumptions, the age-at-maturity of prey is not
necessarily the same as the one of predators. Similar considerations would hold
for a
size-structured population.
We show how to obtain a system of neutral equations with constant
delays from the age structure. In this framework it is possible to
find few heuristically introduced and previously published models, e.g., those
in
\cite{Bartlett1957, Hastings1983} but also examples which are inconsistent with
the biology.
Results on the qualitative behaviour of solutions are provided in Section
\ref{sec:analysis_modelA} Further we study a delay extension of
\eqref{sys:rosenzweig}, which is an example for the class of equations
introduced in Section \ref{sec:model} We shall point out the sensitive
dependence of the model dynamics on the delay and provide numerical examples.\\
\ \\
Throughout this paper, the indices 1 and 2 indicate variables and parameters
related to juvenile and adult individuals, respectively.

%% file: model.tex
\section{Model derivation} 
\label{sec:model}
Before discussing age-structured predator-prey interactions, we briefly
recall the dynamics of an isolated population structured by age. 
Let $n(a,t)$ be the population density with respect to the age $a$ at time $t$.
Biological interpretation suggests that $ \lim \limits_{a \to
\infty} n(a,t) =0$. The classical representation of an isolated population
structured by age is the Lotka-Sharpe model \cite{Sharpe1911},  
\begin{equation}
\label{sys:lotkasharpe}
\begin{aligned}
\frac{\partial}{\partial t}n(a,t) + \frac{\partial}{\partial a}n(a,t) &=-
\mu(a) n(a,t),\\
n(0,t) &= \int_0^\infty b(a) n(a,t)\, da, \\
n(a,0) &= n_0(a).
\end{aligned} 
\end{equation}
Here $\mu:[0,\infty) \rightarrow [0,\infty)$ and $b:[0,\infty) \rightarrow
[0,\infty)$ denote the age-dependent mortality and fertility rate, respectively.
The number of newborns at time $t$ is $B(t) = n(0,t)$. The
continuous function $n_0:[0,\infty) \rightarrow [0,\infty)$ provides the
initial age distribution. With the method of characteristics one finds the
explicit solution of \eqref{sys:lotkasharpe},
\begin{equation}
n(a,t) = \begin{cases}
n_0(a-t) e^{-\int_{a-t}^a \mu(s)\, ds},& \quad  a>t,\\
B(t-a) e^{-\int_{0}^a \mu(s)\, ds}, & \quad a\leq t.
\end{cases}
\label{sys:lotkasharpe_explsolux}
\end{equation}

\noindent The Lotka-Sharpe model \eqref{sys:lotkasharpe} is our point of
departure. We follow \cite{Bocharov2000} and introduce a threshold age,
$\tau>0$, to distinguish juvenile individuals ($a<\tau$) from adult ones
($a>\tau$),
\begin{equation*}
n(a,t) = \begin{cases}
n_1(a,t), & \quad  a \in [0,\tau),\\
n_2(a,t), & \quad a \in (\tau,\infty).
\end{cases}
\end{equation*}
The total juvenile and adult populations at time $t\geq 0$ are thus,
respectively,
\begin{equation*}
\label{def:total_adultjuvenile_popul}
n_1(t)=\int_0^{\tau}n_1(a,t)\,da \qquad \mbox{ and }\qquad
n_2(t)=\int_{\tau}^\infty n_2(a,t)\,da.
\end{equation*}
In the next section we extend this idea to model predator-prey interactions.
Table \ref{tab:notation2} provides an overview of parameters and variables used.

\subsection{Age-structured predator population}
\label{sec:model_agestr_pred}
\begin{figure}
\centering
\includegraphics[width=0.93\textwidth]{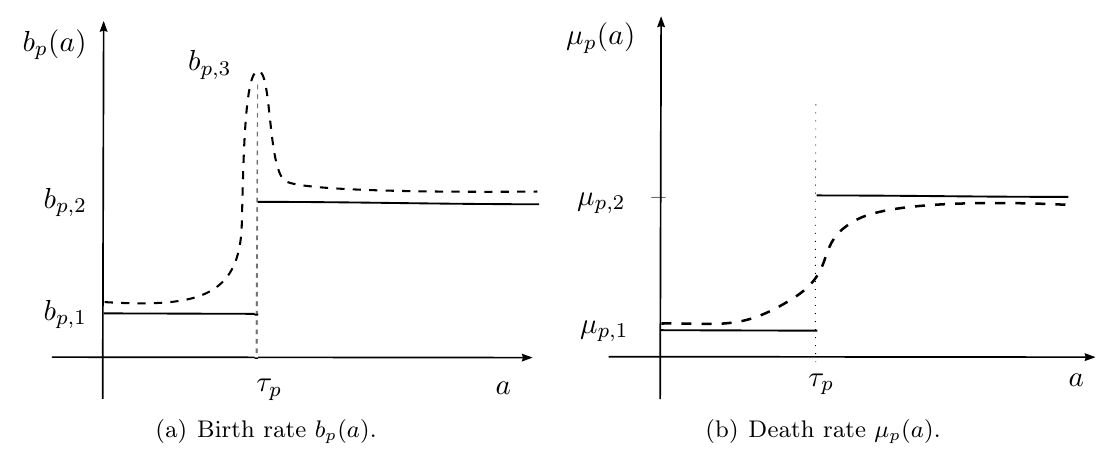}
\caption{Birth and death rates of predators are functions of the age $a$. Solid
lines reflect the model assumptions, dashed
curves represent biologically realistic smooth functions.}
 \label{Fig:pred_birth+death_age_dependent}
 \end{figure}

We consider a predator-prey model where only the predator
population is structured by age. Let $p(a,t)$ denote the predator population
density of individuals of age \(a\) at time \(t\) and, as done above,
distinguish juvenile predators, $p_1(a,t)=p(t,a),\; a \in [0,\taup)$, from
adult ones, $p_2(a,t)=p(t,a)$, $a \in (\taup,\infty)$. Transition from the
juvenile class to the adult one occurs at age $\taup>0$, the
age-at-maturity of the predators. Let \(\taup^+ := \inf\{s : s>\taup\}\)
and \(\taup^- := \sup\{s: s< \taup\}\). The total number of predators, \(P(t)\),
is given by
\begin{equation*}
P(t) \,=\, \int_{0}^{\infty} p(a,t)\,da\, =\,\int_{0}^{\taup^-}p_1(a,t)\,da
+\int_{\taup^+}^{\infty} p_2 (a,t)\,da\,=:\,P_1(t) + P_2(t).
\end{equation*}
Let \(R(t)\) denote the number of prey at time \(t\). When we want to model a
general (non-structured) prey population interacting with juvenile and
adult predators, we expect to work with a system of three equations,
\begin{equation}
\begin{aligned}
\mbox{Prey}\qquad R' &= \alpha(R)R -g(R,
P_1,P_2)\delta(P_1,P_2),\\
\mbox{Juvenile predators}\qquad P_1' &= \pi_1(R, P_1,P_2) - m_1(R, P_1,P_2)
-\gamma_1(P_1,P_2)P_1,\\
\mbox{Adult predators}\qquad P_2' &= \pi_2(R, P_1,P_2)
-\gamma_2(P_1,P_2)P_2.
\end{aligned}
\label{sys:predpreygeneral3_matur}
\end{equation} 
Death of prey is due to
encounters with juvenile and/or adult predators ($\delta(P_1,P_2)$). The terms
$\pi_1 (\pi_2)$ and
$m_1$ describe, respectively, the recruitment into the juvenile (adult) class
and the maturation
from the juvenile class into the adult one. Indeed, in general these processes
can be regulated
by both predator classes. For the dynamics of an isolated population, a similar
system has been introduced in \cite{Nisbet1983}. Recruitment into the juvenile
class is mostly given
by a birth function, whereas into the adult class it occurs by maturation only.
In
\cite{Nisbet1983} it was shown that $m_1(R, P_1,P_2)=\pi_2(R, P_1,P_2)$ is
actually a function of $P_2(t-\taup)$. As in \eqref{sys:predpreygeneral}, we
include the
function $h$ into \eqref{sys:predpreygeneral3_matur} and obtain
\begin{equation}
\begin{aligned}
R' &= \alpha(R)R -g(R, P_1,P_2)\delta(P_1,P_2),\\
P_1' &= \underbrace{h(R, P_1,P_2)\beta_1(P_1,P_2)}_{\pi_1}
- m_1(R, P_1,P_2) - \gamma_1(P_1,P_2)P_1, \\
P_2' &= \pi_2(R, P_1,P_2) -\gamma_2(P_1,P_2)P_2.\\
\end{aligned}
\label{sys:predpreygeneral3}
\end{equation}
It is not straightforward to formulate correctly the above model. In particular,
it might be not clear how to write the terms $m_1$ and $\pi_2$. The approach we
show below provides a physically correct formulation of the mathematical
model. For compactness of notation, in the following
we denote $g(R(t), P_1(t),P_2(t))$ and $h(R(t), P_1(t),P_2(t))$ by $g(t)$ and
$h(t)$,
respectively.\\
\ \\
For the age-structured predator population, we choose birth and death rates in
the form  
\begin{equation*}
\label{def:a-s-pred_birthdeathcoeff}
 \begin{aligned}
\mu_p(a) &= \mupj + (\mupa - \mupj) H_{\taup}(a),\\ 
b_p(a) &= \bpj + (\bpa - \bpj) H_{\taup}(a) +\bpp \delta_{\taup}(a),  
 \end{aligned}
\end{equation*}
where \(H_{\taup} (a)\) is the Heaviside function with a jump at $a=\taup$.
The coefficients $b_{p,k}, \mu_{p,k}, \, k=\{1,2 \},$ represent birth and death
rates of juveniles ($k=1$) and adults ($k=2$). When individuals reach sexual
maturity, at age $a=\taup$, there may be a peak of weight
$\bpp\geq0$ in the fertility rate ($\delta_{\taup}(a)$ is the delta
distribution). A similar
assumption was used in \cite{Bocharov2000}. Figure
\ref{Fig:pred_birth+death_age_dependent} shows the rates $b_p(a)$ and
$\mu_p(a)$.

\noindent Taking into account effects of predation, we put up a modified
Lotka-Sharpe model for $p_1(a,t)$:
\begin{equation*}
\label{sys:p_1_lotkasharpe}
\begin{aligned}
\frac{\partial}{\partial t}p_1(a,t) + \frac{\partial}{\partial a}p_1(a,t) &=- 
\mu_p(a) p_1(a,t) ,\\
p_1(0,t) &= \int_0^\infty b_p(a) p(a,t) da \\
& = \bigl(\bpj P_1(t) + \bpa P_2(t) + \bpp
p_1(\taup^{-},t)\bigr)h(t),\\
p_1(a,0) &= p_1^0 (a),
\end{aligned} 
\end{equation*}
with $p_1^0 (a)\geq 0$ for all  $a \in [0,\taup)$. 
Assuming that no individual dies at the very moment when it becomes adult, $
p_2(\taup^{+},t) =
p_1(\taup^{-},t)$, and that $\lim \limits_{a\to \infty} p_2(a,t)=0$, we have
a similar system for $p_2(a,t)$ with initial age distribution
$p_2^0 (a)\geq 0$ for all 
$a>\taup$.\\
\ \\
The total number of juvenile individuals satisfies
\begin{align*}
P_1'(t) &= \int_{0}^{\taup^{-}}\frac{\partial p_1}{\partial t}(a,t)\,da\,=\,
\int_{0}^{\taup^{-}}-\frac{\partial p_1}{\partial a}(a,t) -\mupj p_1(a,t) da\\
&= p_1(0,t) - p_1(\taup^{-}, t) - \mupj P_1(t), 
\end{align*}
and for the adult population we have
\begin{equation}
P_2'(t) = \int_{\taup{+}}^{\infty}\frac{\partial p_2}{\partial t}(a,t)\,da
= p_1(\taup^{-},t) - \mupa P_2(t).
\label{eq:dotP2} 
\end{equation}
With the explicit solution \eqref{sys:lotkasharpe_explsolux} of a Lotka-Sharpe
model, we find for $t <\taup$
\begin{align*}
p_1(\taup^{-},t) = p_1^0 (\taup -t) e^{-\mupj t},
\end{align*}
and for $t \geq \taup$, 
\begin{equation*}
p_1(\taup^{-},t) = \Bigl[\bpj h(t-\taup)P_1(t-\taup) +
\bpa h(t-\taup)P_2(t-\taup) + \bpp h(t-\taup)  p_1(\taup^{-},t-\taup)\Bigr]
e^{-\mupj\taup}. 
\end{equation*} 
With \eqref{eq:dotP2}, for $t \geq \taup$ we have
\begin{equation*}
\begin{aligned}
p_1(\taup^{-},t)& = \Bigl[\bpj h(t-\taup)P_1(t-\taup) +
\bpa h(t-\taup)P_2(t-\taup)\\ 
&\phantom{=} + \bpp h(t-\tau_p)
\Bigl(P_2'(t-\taup)+\mupa
P_2(t-\taup)\Bigr) \Bigr]e^{-\mupj \taup}. 
\end{aligned}
\end{equation*}
For the prey population, we have in general the first equation in
\eqref{sys:predpreygeneral3}.
One possible choice for $\alpha(R)$ could be the logistic growth with carrying
capacity
\(K > 0\) and net growth rate \(r > 0\),
\begin{equation}
\label{model1_prey}
R'(t) = \underbrace{r \left(1-\frac{R(t)}{K}\right)}_{\alpha(R(t))}R(t) -
g(t)\delta(P_1(t),P_2(t)).
\end{equation}
For \(t < \taup\) the predator population satisfies
\begin{equation}
\begin{aligned}
P_1'(t) &= \bigl(\bpj h(t)- \mupj\bigr) P_1(t) +  \bpa h(t) P_2(t)
+\bigl(\bpp h(t)-1\bigr) p_1^0(\taup-t)e^{-\mupj t}, \\
P_2'(t) &=  - \mupa P_2(t) + p_1^0(\taup-t)e^{-\mupj t}.
\end{aligned}
\label{model1a}
\end{equation}
For \(t\geq \taup\) we have 
\begin{equation}
\begin{aligned}
P_1'(t) &= \bigl(\bpj h(t) -\mupj\bigr) P_1(t) + \bpa h(t)P_2(t)
+ \bigl(\bpp h(t) -1\bigr) \bpj h(t-\taup) P_1(t-\taup) e^{-\mupj \taup}\\
&\phantom{=}  + \bigl(\bpp h(t)-1\bigr) \Bigl(\bpa h(t-\taup)+ \bpp
h(t-\taup) \mupa\Bigr) e^{-\mupj \taup} P_2(t-\taup)\\
&\phantom{=}  + \bigl(\bpp  h(t)-1\bigr) \bpp h(t-\taup) e^{- \mupj \taup}
P_2'(t-\taup),\\[0.3em]
P_2'(t) &= \Bigl[\bpj h(t-\taup) P_1(t-\taup) +\Bigl(\bpa
h(t-\taup)+ \bpp h(t-\taup)\mupa\Bigr) P_2(t-\taup)\\ 
&\phantom{= \Bigl[}  + \bpp h(t-\taup)P_2'(t-\taup)\Bigr]e^{-\mupj \taup} -
\mupa P_2(t).
\end{aligned}
\label{model1b}
\end{equation}
\noindent We have obtained a class of systems \eqref{model1_prey},
\eqref{model1a}, respectively \eqref{model1_prey}, \eqref{model1b}, in which we
can cast several examples from the literature.\\
\indent Consider the system for $t\geq \taup$. For $\bpp=0$, we have a system
of the form \eqref{sys:predpreygeneral3}. It is now clear how $m_1$
and $\pi_2$ should be formulated in terms of $h(t-\taup)$ and
$P_j(t-\taup),\,j=1,2$. Assuming that juvenile individuals
are not fertile ($\bpj=0$), we obtain 
\begin{equation}
\begin{aligned}
R'(t) & =r \left(1-\frac{R(t)}{K}\right)R(t) -g(t)\delta(P_1(t),P_2(t)),\\
P_1'(t) &= -\mupj P_1(t) + \bpa h(t)P_2(t)- \bpa h(t-\taup)e^{-\mupj \taup}
P_2(t-\taup),\\
P_2'(t) &= \bpa h(t-\taup)P_2(t-\taup)e^{-\mupj \taup} - \mupa P_2(t).
\end{aligned}
\label{model1b_example}
\end{equation}
A proper choice of birth and death rates and of the functions $h$ and $g$
yields,
e.g., the model by Gourley and Kuang \cite{Gourley2004}. Another system of the
form
\eqref{model1_prey}, \eqref{model1a} is the third model in \cite{Nunney1985a},
though there is here no equation for $P_1$ (predators are meant to be adult
predators). A ``complementary'' model to \eqref{model1_prey}, \eqref{model1a} is
provided in \cite{Hastings1983}, where maturation is considered only for the
prey population. In addition our approach shows that sometimes
examples from the literature are 
inconsistent with the biological phenomena. For example Ross \cite{Ross1972}
suggests a model in
which the recruitment into the adult predator population is given by $\bpa
R(t)P_2(t-\taup)$. Apparently, the delay in the prey density has been neglected.

\subsection{Age-structured predator and prey population} 
\label{sec:model_agestr_predprey}
Now we assume that the prey population is structured by age, too. As in Section
\ref{sec:model_agestr_pred}, we
simplify the age structure by introducing an age-at-maturity, $\tau_r>0$, for
the prey. Again, we
shall consider only juvenile and adult individuals.
The advantage of an age structure for the prey population is not only that we
have age-dependent fertility and mortality rate, but we can also assume that
predation depends on the age of the prey. This allows us to include
age-specific predation into the model and to reflect different
possible settings from biology. In general, a mathematical model which includes
juvenile and
adult predators (respectively, prey) is an extension of system
\eqref{sys:predpreygeneral3} and
can be described by a system of four equations
\begin{equation*}
\begin{aligned}
R_1' &= \alpha_1(R_1,R_2)a_1(R_1,R_2) -q_1(R_1,R_2,P_1,P_2)
-g_1(R_1,R_2, P_1,P_2)\delta_1(P_1,P_2),\\
R_2' &= \rho_2(R_1,R_2,P_1,P_2) -g_2(R_1,R_2,P_1,P_2)\delta_{2}(P_1,P_2),\\
P_1' &= h(R_1,R_2, P_1,P_2)\beta_1(P_1,P_2) - m_1(R_1,R_2, P_1,P_2) -
\gamma_1(P_1,P_2)P_1, \\
P_2' &= \pi_2(R_1,R_2, P_1,P_2)-\gamma_2(P_1,P_2)P_2.\\
\end{aligned}
\end{equation*}
Here the term $q_1$ indicates maturation of juvenile prey and $\rho_2$ the
recruitment into the
adult prey class. As in \eqref{sys:predpreygeneral3} predation and recruitment
rates can be due to all
classes of individuals. For simplicity we
denote the functions $g_j(\cdot)$ and
$h(\cdot)$ by $g_j(t)$, $j=1,2$, and $h(t)$, respectively.\\
\ \\ 
As in Section \ref{sec:model_agestr_pred} we start with age-structured prey and
predator populations and
derive DDEs for juvenile and adult individuals. For the predator population we
have
essentially the systems \eqref{model1a} and \eqref{model1b}. Equations for the
prey
population can be obtained in a similar way, however, we have to take into
account age-specific predation. We assume that only adult preys
undergo predation, that is $g_1(t)\equiv 0$. This assumption
fits several insect species, where individuals in the egg and larval stage are
well protected
\cite{Williams1995}. It follows that $q_1(R_1,R_2,P_1,P_2)=q_1(R_1,R_2)$ and
$\rho_2(R_1,R_2,P_1,P_2)=\rho_2(R_1,R_2)$.\\
\indent For simplicity of computation, let $g_2(t)=\mu_{r,2}R_2(t)$ (however a
different choice of
$g_2$ is also possible, cf. Section \ref{sec:intro}). Under the assumption that
juveniles die
at constant
rate $\murj$, we have
\begin{align*}
\mu_r(a) &= \mu_{r,1} + \bigl(\delta_2(P_1(t), P_2(t))\mu_{r,2} -
\mu_{r,1}\bigr)
H_{\tau}(a).
\end{align*}
Let $r_1(a,t),\, a \in [0,\tau_r)$ be the density of juvenile prey at time
$t\geq 0$. Analogously, $r_2(a,t),\, a \in (\tau_r, \infty)$ denotes the
density of adult prey  at time $t\geq 0$. Let
$R_1(t)$ (respectively, $R_2(t)$) be the total juvenile (respectively, adult)
population size at time $t$.\\
As in the previous section, we consider the PDE model for  $r_1(a,t)$ and
$r_2(a,t)$ and obtain a system of differential equations for $R_1$ and $R_2$.
Computation yields for \(t<\tau_r\)
\begin{equation}
\label{model2a_j_a_prey}
\begin{aligned}
R_1'(t)&= \bigl(b_{r,1}-\mu_{r,1}\bigr)R_1(t) +
b_{r,2}R_2(t)+\bigl(b_{r,3}-1\bigr) r_1^0 (\tau_r -t) e^{-\mu_{r,1} t},\\
R_2'(t)&=  -\mu_{r,2}\delta_2(P_1(t), P_2(t))R_2(t) +  r_1^0 (\tau_r -t)
e^{-\mu_{r,1} t},
\end{aligned}
\end{equation}
and for \(t \geq \tau_r\)
\begin{equation}
\label{model2b_j_a_prey}
\begin{aligned}
R_1'(t)&= \bigl(b_{r,1}-\mu_{r,1}\bigr)R_1(t) + b_{r,2}R_2(t) + \bigl(b_{r,3} 
-1\bigr) b_1 R_1(t-\tau_r) e^{-\mu_{r,1} \tau_r}\\
&\phantom{=}+ \bigl(b_{r,3} -1\bigr) \Bigl(b_{r,2}+ b_{r,3} 
\mu_{r,2}\Bigr) e^{-\mu_{r,1} \tau_r}R_2(t-\tau_r)\\
&\phantom{=}+ \bigl(b_{r,3}  -1\bigr) b_{r,3}e^{- \mu_{r,1} \tau_r}
R_2'(t-\tau_r),\\[0.3em]
R_2'(t) &= \Bigl[b_{r,1} R_1(t-\tau_r) +\Bigl(b_{r,2}+ b_{r,3}\mu_{r,2}\Bigr)
R_2(t-\tau_r)\\ 
&\phantom{=}+ b_{r,3}R_2'(t-\tau_r)\Bigr]e^{-\mu_{r,1} \tau_r} - \mu_{r,2}
\delta_2(P_1(t), P_2(t))R_2(t).
\end{aligned}
\end{equation}
\noindent Hence we have obtained a general class of predator-prey systems in
which we can cast
several examples from the literature (of course a proper choice of birth and
death rates and of the
functions $h$ and $\delta_2$ is necessary). As for the systems
\eqref{sys:predpreygeneral} and
\eqref{sys:predpreygeneral3}, also in this case there is a certain relation
between predation
response and fecundity response of predators. It is meaningful to choose $h(t)$
proportional to a
linear combination of $g_1(t)$ and $g_2(t)$, that is, the growth rate of
predators depends on the
(eaten) prey (cf. \cite{Nunney1985a, Kuang1991, Beretta2002}). If predation is
only
due to adult predators, then we have $\delta_j(P_1,P_2)=\delta_j(P_2),\,
j=1,2$.\\
\indent It is worth noticing that often in previous models no distinction
between
juvenile and adult individuals has been made (cf. \cite{Bartlett1957,
Nunney1985a, Kuang1991}). A
comparison with our model suggests that, mostly, heuristically
introduced equations describe the dynamics of adult prey and predators. A
special case, in which
predators and prey species have the same maturation time was
given in \cite{Nunney1985a}. Further, when two
populations of juveniles and adult individuals are explicitly introduced, e.g.,
in
\cite{Gourley2004}, then usually it is assumed that juvenile individuals are not
fertile, that
is, $b_1=0$ as in \eqref{model1b_example}.\\
\indent Our approach guarantees that the obtained mathematical model is
consistent with biology. This is not the case for all models. For example
in \cite{Beretta2002} death of prey is described by the term
$-\mu_rP(t)R(t-\sigma)$, whereas
recruitment into predator population is given by $bP(t-\tau)R(t)$. The two
delays $\sigma$ and
$\tau$ seem to have no connection with maturation times and it remains unclear
how they should be motivated from a biological point of view.\\
\ \\
In general, prey and predators are not characterized by the same
age-at-maturity, that is, 
we could have \(\tau_p \neq \tau_r\). Hence, there are three time intervals
which we
have to consider separately. Let \(\tau_{min} := \min\{\tau_p, \tau_r\}\)
and \(\tau_{max} := \max\{\tau_p, \tau_r\}\). For $t\in [0,\tau_{min})$, we have
a nonautonomous ODE system \eqref{model1a}, \eqref{model2a_j_a_prey}, whose
right-hand side depends on the initial age distribution of the underlying PDE
model. For $t\in [\tau_{min}, \tau_{max})$, we have a system
consisting of two nonautonomous ODEs and two neutral equations
with constant delay. If $\tau_{min} = \tau_p$, the system is given by 
\eqref{model1b} -- \eqref{model2a_j_a_prey}, whereas it
is given by \eqref{model1a}, \eqref{model2b_j_a_prey}, if $\tau_{min} = \tau_r$.
In both cases, we
have a
combination of initial data given by the initial age distribution of the PDE
model and the solution of the system on \([0,\tau_{min})\). For $t \in
[\tau_{max}, \infty)$, we have a system of four
neutral equations with two constant delays
\eqref{model1b}, \eqref{model2b_j_a_prey}. Here the initial data is given by
the solution on \([0, \tau_{max})\).\\
\ \\
In the general settings presented above, we have assumed that there is a peak
of weight $\bpp$ in the birth rate when individuals reach maturity. Such a peak
in the fertility rate was observed, e.g., in loggerhead turtles
\cite{Crouse1987}, as well as in many insect populations \cite{Williams1995}.
From a mathematical point of view, $\bpp \neq 0$ yields a neutral
equation. To the best of our knowledge,
there are only few other examples of neutral equations for predator-prey
interactions.  Kuang \cite{Kuang1991} suggests a system with neutral term in the
equation for the prey,
\begin{equation}
\label{model_prey_kuang}
R'(t) = r R(t) \left(1-\frac{R(t-\tau_r)+\dot R(t-\tau_r)}{K}\right) -
P(t)\rho(R(t)).
\end{equation}
This model is essentially obtained by extending the ``neutral logistic
equation'' in \cite{Gopalsamy1988} with a further death term due to predation.
The formal derivation of a DDE from the age structure yields in 
\eqref{model2b_j_a_prey} a delay in the recruitment term of the prey, in
contrast to
equation \eqref{model_prey_kuang}, where the delay appears in the death term
(see also
\cite{Bocharov2000} for similar observations for single population
dynamics).\\
\indent However, in many  species, one does usually not observe a peak in the
fertility rate, but rather a jump, when individuals become sexually mature,
see e.g. the data set in \cite{Novoseltsev2003}. In the following we
shall neglect the peak in $b(a)$, both for the prey and for the predator
population, i.e., $b_{p,3} = 0= b_{r,3}$. This assumption reduces the
problem to a system of (non-neutral) equations with constant delays.
\begin{table}[h]
\centering
\begin{tabular}{l|l}
\hline
\textbf{Symbol} & \textbf{Description}\\
\hline
\hline
\(p(a,t)\) & density of predators of age \(a\) at time \(t\)\\
\(p_j(a,t)\) & density of juvenile/adult predators of age \(a\) at time \(t\)\\ 
\(P(t)\) & size of predator population at time \(t\)\\
\(P_j(t)\) & size of juvenile/adult predator population at time \(t\)\\
\(r(a,t)\)& density of prey of age \(a\) at time \(t\)\\
\(r_j(a,t)\) & density of juvenile/adult prey of age $a$ at time $t$\\
\(R(t)\) & size of prey population at time \(t\)\\
\(R_j(t)\) & size of juvenile/adult prey population at time \(t\)\\
\(\tau_p\) & age at maturity of predator\\
\(\tau_r\) & age at maturity of prey\\
\(\mu_{p,j}\) &  death rate of juvenile/adult predator\\
\(b_{p,j}\) & fertility rate of juvenile/adult predator\\
\(b_{p,3}\) &fertility rate of predator at age \(\tau_p\)\\
\(\mu_{r,j}\) & death rate of juvenile/adult prey\\
\(b_{r,j}\) & fertility rate of juvenile/adult prey\\
\(b_{r,3}\) &fertility rate of prey at age \(\tau_r\)\\
\hline
\end{tabular}
\caption{Variables and parameters for the models in Section
\ref{sec:model}. The indices $j=1$ and $j=2$ indicate juvenile and adult
individuals, respectively. All quantities are nonnegative.}
\label{tab:notation2}
\end{table}

%% file: analysis_modelchanged.tex
\section{Analytical Results}
\label{sec:analysis_modelA}
Here we provide some analytical results for the model with one delay. Much more
challenging is the
case of two (or several) delays, which shall be studied in a forthcoming
paper.\\
\ \\
Consider the model \eqref{model1b_example} for $t\geq \taup$, and let
$g(t)=g(R(t), P_2(t))$ and
$h(t)=h(R(t), P_2(t))$. For simplicity of notation, we omit the $p$-index in the
coefficients and in
the delay. For $t< \tau$ we have
\begin{align}
R'(t) & =r \left(1-\frac{R(t)}{K}\right)R(t)
-g(t)\delta(P_1(t),P_2(t)),\label{model1ef_prey}\\
P_1'(t) &= - \mu_{1} P_1(t) +  b_2 h(t) P_2(t) -p_1^0(\tau-t)e^{-\mu_1 t},
\label{model1e1}\\
P_2'(t) &=  - \mu_2 P_2(t) + p_1^0(\tau-t)e^{-\mu_1 t}. \label{model1e2}
\end{align}
Taken together the model, which describes the dynamics in the whole interval
$[0,\infty)$ can be
seen as a general formulation of the system by Gourley and Kuang
\cite{Gourley2004}.\\
\ \\
\noindent Preservation of positivity is a crucial factor in mathematical
biology. However, systems
of DDEs can possibly have negative solutions even when the initial
functions are nonnegative \cite{Smith2011}. In the following we provide criteria
to guarantee
nonnegative solutions of \eqref{model1b_example}.\\
\indent The right-hand side of \eqref{model1ef_prey}--\eqref{model1e2} depends
on the initial age
distribution $p^0_1(a)$ of a PDE system of the form \eqref{sys:lotkasharpe} for
the juvenile
predator. If $p^0_1(a)\geq 0,\,a\geq0$ is known, we take the solution of 
\eqref{model1ef_prey}--\eqref{model1e2} as history function for
\eqref{model1b_example} and we obtain nonnegative solutions.
Now consider the case in which $p^0_1(a)$ is not known. Following
\cite{Bocharov2000}, we start with
(\ref{model1e2}) and find
\begin{align*}
\frac{d}{d t}\bigl(P_2(t) e^{\mu_2 t}\bigr) e^{(\mu_1-\mu_2)t} &= p_1^0 (\tau
-t) \qquad (\geq 0).
\end{align*}
That is, the function \(P_2(t) e^{\mu_2 t}\) is nondecreasing
on \([0,\tau]\). Integration in \([0,\tau]\) yields
\begin{align}
P_1(0) = P_2(\tau) e^{\mu_1\tau} - P_2(0) - (\mu_1 -\mu_2)\int_{0}^{\tau} P_2(t)
e^{\mu_1 t}  dt. \label{57}
\end{align}
We use \eqref{model1e1} to substitute the term \(p_1^0 (\tau -t) e^{-\mu_1 t}\)
and obtain
\begin{align*}
P_1'(t) &= -\mu_1P_1(t) + \bigl(b_2h(t)-\mu_2\bigr)P_2(t) -P_2'(t).
\end{align*}
The last equation can be solved with variation of constants
and integration by parts,
\begin{equation}
\begin{aligned}
P_1(t) & = \bigl(P_1(0) + P_2(0)\bigr)e^{-\mu_1 t} - P_2(t) + \int_{0}^{t}
\bigl(b_2 h(s) -\mu_2\bigr)P_2(s)e^{\mu_1 (s-t)}ds\\ 
&\phantom{=}\; + \mu_1 \int_{0}^{t} P_2(s) e^{\mu_1 (s-t)}ds.
\end{aligned} 
\label{59}
\end{equation}
As the equation \eqref{model1ef_prey} is an ODE, there
is no problem with positivity of solutions, as long as we choose a nonnegative
initial value \(R(0)\). The solution exists and is unique, given 
continuously differentiable \(g\) and $\delta$.\\
\indent We define an operator \(T: (C[0,\tau])^2 \rightarrow C[0,\tau]\) so that
\((T(R, P_2))(t) = P_1(t)\) for \(t \in [0,\tau]\) and \(P_1(t)\) is given by 
(\ref{59}) with \(P_1(0)\) as in (\ref{57}). The cone \(\mathcal{K}\) is defined
by
\begin{align*}
\mathcal{K} := &\Bigl\{ (R, P_1,P_2) \in (C[0,\tau])^3 :\;P_2(0) \geq 0,\\
& \phantom{= =} P_2(t) e^{\mu_2 t} \text{ is nondecreasing in } [0,\tau],\;P_1=
T(R, P_2), R(0) \geq 0 \Bigr\}. 
\end{align*}

\begin{proposition}
Let a PDE system of the form
\eqref{sys:lotkasharpe} for $p_1(a,t)$ be given with initial age distribution
$p_1^0(a) \geq 0$.
Then the functions $R(t), P_1(t), P_2(t)$ defined in Section
\ref{sec:model_agestr_pred}
satisfy system \eqref{model1ef_prey}--\eqref{model1e2} and
$(R,P_1,P_2)\big|_{[0,\tau]}\in \mathcal{K}$.\\
\ \\
Conversely, for $(\tilde{R}, \tilde{P}_1,\tilde{P}_2) \in
\mathcal{K}$, there is an initial age distribution \(p_1^0(a) \geq 0\) such that
the
solution \((R, P_1,P_1)\) of the system \eqref{model1ef_prey}--\eqref{model1e2}
satisfies
$(R,P_1,P_2)\big|_{[0,\tau]} =(\tilde{R}, \tilde{P}_1,\tilde{P}_2)$.
\end{proposition}

\noindent \textbf{Proof.}
The first part has already been shown. The rest of the proof follows similar to
\cite{Bocharov2000}, with the difference that here the operator $T$ has two
arguments.

\subsection{Linearised stability}
\subsubsection{The case $\tau=0$} 
First we consider \eqref{model1b_example} with \(\tau = 0\). 
From a biological point of view, this means that newborn individuals are
sexually mature.
The juvenile class $P_1$ loses its original meaning. Evidently,
$\lim \limits_{t\to
\infty} P_1(t)=0$. Fixed points of the system are the trivial
equilibrium \(E_0 =
(0,0,0)\), in which we find neither prey nor predator, the point
\(E_R = (K,0,0)\), where only prey are present and a \textit{coexistence
equilibrium} 
\(E_C = (R^*,0,P_2^*)\), with $R^* \in (0,K)$ and $P_2^* > 0$. Such an
equilibrium is
obtained as the intersection of the curve
\(b_2h(R^*, P_2^*) = \mu_2\) and the prey isocline $rR^*(1-\frac{R^*}{K})=g(R^*,
P_2^*)\delta(0,P_2^*)$ in the positive cone. The number of coexistence points
depends on the choice
of the functions \(h\) and \(g\).\\
\ \\
We denote by $g_R$ and $g_{P_2}$ the derivative
of $g$ with respect to $R$ and ${P_2}$, respectively (analogously for $h$).
The biology suggests that the functions $h$ and $g$ are
nondecreasing both in $P_2$
and $R$, hence all the partial derivatives are nonnegative. This assumption
includes the possibility that, when the
prey population has reached its carrying capacity, the prey dies for reasons
different than predation, e.g.,  external harvesting or migration.
However, we assume that $g_R(R,0)= 0$, for all
$R\geq 0$, i.e., in
absence of predators no prey dies. The function $\delta(P_1,P_2)$ is
nondecreasing in both components.

\begin{proposition}
The trivial equilibrium \(E_0=(0,0,0)\) is always
unstable.\\
The point \(E_R = (K,0,0)\) is locally asymptotically stable if $\mu_2>
b_2h(K,0)$.
\end{proposition}

\noindent \textbf{Proof.}
The proof follows with standard linear stability theory. Computation can be
found in \cite{Mohr2012}.\\
\ \\
\noindent Consider the coexistence equilibrium $E_C = (R^*, 0,
P_2^*)$. As the $P_1$-direction is always stable, it is sufficient to
investigate the matrix $A=\left( a_{l,k}
\right)_{l,k=1,2}$,
\begin{displaymath}
A
= \begin{pmatrix}
r- 2 \frac{r}{K}R^* -g_R(R^*, P_2^*)\delta(0,P_2^*)
& - g_{P_2}(R^*, P_2^*)\delta(0, P_2^*) - g(R^*, P_2^*)\delta_{P_2}(0, P_2^*)\\
& \\
b_2h_R(R^*, P_2^*)P_2^* &   b_2h_{P_2}(R^*, P_2^*)P_2^*
\end{pmatrix}.
\end{displaymath}
Notice that $a_{1,2}\leq 0$, whereas $a_{2,1}$ and $a_{2,2}$ are
nonnegative. The coexistence point is stable when $\det(A)>0$ and
$\tr(A)<0$.

\subsubsection{The case $\tau>0$} 
\label{sec:ana_del_case}
As in the case $\tau=0$ we have the equilibria $E_0
=(0,0,0)$ and $E_R=(K,0,0)$. Coexistence equilibria depend on the choice of
$h$ and $g$. In Section \ref{sec:example_rosen} we show an example in which the
coexistence point
$E_C=(R^*,P_1^*,P_2^*)$ is uniquely determined. We say a
coexistence point is \textit{feasible} when it is in the positive cone
and all its components are bounded.\\
\ \\
\noindent Linearised stability analysis \cite{Smith2011} allows to reduce
the problem to a standard linear equation of the form $x'(t) = Ax(t) +
Bx(t-\tau)$. It is immediate to prove that the delay has no
influence on the stability of the trivial equilibrium $E_0$. It is locally
asymptotically stable if \(r < g_R(0,0)\delta(0,0)\) holds, otherwise unstable
\cite{Mohr2012}.\\
\indent Linearisation about the point $E_R$ yields the characteristic
equation 
\begin{align*}
\bigl(\lambda + r +g_R(K,0)\delta(0,0)\bigr) \bigl(\lambda +\mu_1\bigl)
\bigl(\lambda
+\mu_2 - e^{-(\mu_1 +\lambda)\tau}b_2h(K,0)\bigr) = 0.
\end{align*}
The characteristic roots are $ \lambda_1 = -r -g_R(K,0)\delta(0,0)\in
\R,\,\lambda_2 = -\mu_1
\in \R$ and \(\lambda_3\) which is given by the equation
\begin{align}
\lambda +\mu_2 - e^{-(\mu_1 +\lambda)\tau}b_2h(K,0) = 0. \label{equa_lambda3}
\end{align}
This equation has the form $\lambda -a-be^{-\tau \lambda} = 0,$
with \(a = -\mu_2\) and \(b =e^{-\mu_1 \tau}b_2 h(K,0) > 0\) for all
\(\tau >0\). The delay in the coefficients of the
characteristic equation can make the stability analysis challenging.
However, Theorem 6 in \cite{Frasson2009} ensures that equation
\eqref{equa_lambda3} has a real dominant root (the condition  $-e^{-1} \leq
b\tau e^{-a\tau}$ holds for all $\tau >0$). Hence it is sufficient to
investigate the real
characteristic roots of
(\ref{equa_lambda3}), which are given by the intersections of the line \(y=
x + \mu_2\) with the curve \(y = b_2 h(K,0) e^{-\mu_1 \tau} e^{-x\tau}\). 
\begin{proposition}
\label{prop:DDE_stabER}
The equilibrium \(E_R\) of system \eqref{model1b_example} is locally
asymptotically stable if it holds that
$$ r+g_R(K,0)\delta(0,0)>0\qquad \mbox{and} \qquad
\mu_2 > b_2 h(K,0) e^{-\mu_1 \tau}.$$
\end{proposition}
\noindent Assume that there is a uniquely determined coexistence point
$E_C=(R^*, P_1^*, P_2^*)$.
For simplicity of computation we assume that predation is only due to
adult predators, i.e., $\delta(P_1,P_2)=\delta(P_2)$.
We are interested in stability properties of $E_C$ with respect to the
delay $\tau>0$. To this purpose we shall make use of results in 
\cite[Sec. 2]{Beretta2002}. Linearisation of \eqref{model1b_example} about
$E_C$ yields the characteristic equation $G(\lambda; \tau) = 0$, with
\begin{align*}
G(\lambda; \tau) &= \bigl(\lambda +\mu_1 \bigr) \Bigl[\Bigl(\lambda +\mu_2-
e^{-(\lambda +\mu_1) \tau} \bigl(b_2 h(R^*,P_2^*) + b_2 h_{P_2}(R^*, P_2^*)
P_2^*\bigr)\Bigr)\\
&\;\phantom{=}\cdot\Bigl(\lambda -r +2 \frac{rR^*}{K} + g_R(R^*, P_2^*)
\delta(P_2^*)\Bigr)\\
 &\phantom{=}+ e^{-(\lambda +\mu_1) \tau} b_2 h_R(R^*,P_2^*)
P_2^*\Bigl( g_{P_2}(R^*, P_2^*)\delta(P_2^*) + g(R^*, P_2^*)
\delta'(P_2^*)
\Bigr)\Bigr].
\end{align*}
Hence we have one real root \(\lambda = -\mu_1\), whereas other roots are
determined by the term in the square brackets. The
above characteristic equation can be written in the form 
\begin{align*}
P(\lambda; \tau) + Q(\lambda; \tau)e^{-\lambda \tau} = 0,
\end{align*}
with
\begin{equation*}
\begin{aligned}
P(\lambda; \tau) & =  \lambda^2 + \left(2\frac{r}{K} R^* +
g_R(R^*,P_2^*)\delta(P_2^*) -r +\mu_2\right)\lambda\\
&\phantom{=} +\left( 2\frac{r}{K} R^*+ g_R(R^*,P_2^*)\delta(P_2^*) - r
\right)\mu_2,\\
Q(\lambda; \tau) &= -e^{-\mu_1\tau}\left(b_2 h(R^*,P_2^*) +
b_2h_{P_2}(R^*,P_2^*)P_2^*\right)\lambda\\
&\phantom{=} -e^{-\mu_1\tau}\biggl[\left(b_2
h_R(R^*,P_2^*)+ b_2h_{P_2}(R^*,P_2^*)P_2^*\right)\left(2\frac{r}{K}R^* +
g_R(R^*,P_2^*)\delta(P_2^*) -r\right)\\
&\phantom{=} - b_2 h_R(R^*, P_2^*)
\left(g_{P_2}(R^*,P_2^*)\delta(P_2^*) + g(R^*,P_2^*) \delta'(P_2^*)\right)
P_2^*\biggr]. 
\end{aligned}
\end{equation*}
Let
\begin{equation}
\label{betagammaeps}
 \begin{aligned}
\beta &= 2\frac{r}{K} R^* + g_R(R^*, P_2^*)\delta(P_2^*) -r, \\
\gamma &= b_2 h(R^*,P_2^*) +
b_2h_{P_2}(R^*,P_2^*)P_2^*,\\
\epsilon & = b_2 h_R(R^*, P_2^*)
\Bigl(g_{P_2}(R^*,P_2^*)\delta(P_2^*) + g(R^*,P_2^*) \delta'(P_2^*)\Bigr)
P_2^*,
\end{aligned}
\end{equation}
and obtain
\begin{displaymath}
 P(\lambda; \tau) = \lambda ^2 + (\beta +\mu_2)\lambda+ \mu_2\beta,\qquad
Q(\lambda; \tau) = -e^{-\mu_1\tau} \gamma \lambda - e^{-\mu_1\tau} \bigl(\gamma
\beta -\epsilon\bigr).
\end{displaymath}
Stability properties of $E_C$ are determined by the real part of the
characteristic roots. Define the functions
\begin{align*}
M: [0,\infty) \to \R,  &\;\; \tau\, \mapsto \, \beta ^2 + \mu_2 ^2 - e^{-2\mu_1
\tau} \gamma ^2, \\
N: [0,\infty) \to \R,  &\;\; \tau\, \mapsto \, e^{-2\mu_1 \tau}(\gamma \beta
-\epsilon)^2- \mu_2 ^2 \beta ^2.
\end{align*}
As the delay appears in the coefficients of the
characteristic equation (notice that $\tau$ is also in the coordinates
of the point
$E_C$) it is difficult to determine precisely the value of the delay at which
stability switches occur. However, we have the following result.
\begin{theorem}
Let $\beta\bigl(\gamma- \mu_2 e^{\mu_1 \tau}\bigr) \neq \epsilon$ with $\tau
\geq 0$. The stability of $E_C$ depends on the value of the delay $\tau \geq 0$
as follows.
\begin{enumerate}
\item[(1)] Assume \(M(\tau) \geq 0\). If \(\tau\) increases and enters
$S_c := \left\{\tau \in \mathbb{R}_{0}: M(\tau) \geq 0,\; N(\tau) >0\right\}$,
stability switches may occur as \(\tau\) increases within $S_c$. If $\tau \notin
S_c$, no stability switch occurs.
\item[(2)] Assume \(M(\tau) < 0\). If \(\tau\) increases and enters
$S_c := \left\{\tau \in \mathbb{R}_0 : M(\tau) < 0,\;  N(\tau) \geq
-\frac{M(\tau)^2}{4}\right\}$, stability switches may occur as \(\tau\)
increases within \(S_c\). If $\tau \notin S_c$, no stability switch occurs.
\end{enumerate}
\end{theorem}
\noindent \textbf{Proof.}
We verify the hypothesis of \cite[Sec. 2]{Beretta2002}. The functions \(P\) and
\(Q\) do not have any common imaginary root $\lambda=\pm iy,\;y \in \R_+$.
Indeed
$$P(iy; \tau)=-y^2+i(\beta+\mu_2)y
+\mu_2\beta\,=\,0\,=\,-e^{-\mu_1\tau}\left(i\gamma y +\gamma \beta
-\epsilon\right)= Q(iy; \tau)$$
only for $y=0$. 
Further we have $\overline{P(-iy, \tau)} = P(iy, \tau)$ and $\overline{Q(-iy,
\tau)} =Q(iy, \tau)$ for $y \in \mathbb{R}$ and for all $\tau\geq 0$.
Observe that
\begin{align*}
P(0;\tau) + Q(0;\tau)=\beta \left(\mu_2 -\gamma e^{-\mu_1 \tau}\right)-\epsilon
e^{-\mu_1 \tau}\neq 0, \qquad \mbox{for all}\, \tau\geq0.
\end{align*}
Some computation (see \cite{Mohr2012}) shows that for all $\tau \geq 0$, 
$$\limsup\{|Q(\lambda, \tau)|/|P(\lambda, \tau)| : |\lambda| \rightarrow \infty,
 \Re(\lambda)\geq 0 \}<1.$$
\noindent Consider purely imaginary roots
\(\lambda = iy\) for \(y \in \mathbb{R}_+\). Then we have
\begin{equation*}
\begin{aligned}
P_R:= \Re(P(iy; \tau)) = \mu_2 \beta -y^2, & \qquad P_I := \Im(P(iy; \tau)) = 
(\beta +\mu_2)y,\\
Q_R := \Re(Q(iy; \tau)) =  - e^{-\mu_1\tau} \bigl(\gamma \beta
-\epsilon\bigr), & \qquad Q_I := \Im(Q(iy; \tau)) = -e^{-\mu_1\tau} \gamma y.
\end{aligned}
\end{equation*}
Hence,
\begin{equation*}
F(y;\tau) \,=\,P_R ^2 +P_I ^2 - Q_R ^2 -Q_I^2 \,=\,y^4 +M(\tau)y^2 -N(\tau).
\end{equation*}
With \(z :=y^2\), the zeros of $F(y;\tau)$ are given by \(y_{1,2} =
\sqrt{z_\pm}\) and \(y_{3,4} = -\sqrt{z_\pm}\), where
\begin{align*}
z_\pm = -\frac{M(\tau)}{2} \pm
\sqrt{\frac{M(\tau)^2}{4} +N(\tau)}. 
\end{align*}
For each $\tau \geq 0$ we have at most four different real roots of $F(y;\tau)$,
hence also the last hypothesis is satisfied. We search for values of \(\tau\)
such that \(F\) has at least one strictly
positive root. If $M(\tau) \geq 0$, $z_-$ is always negative or complex, but
$z_+ >0$ if and only if $N(\tau) >0$. If $M(\tau) <0$, $z_- >0$ if and only of
$N(\tau) >0$ and $z_+ >0$ if and only if $N(\tau) \geq -\frac{M(\tau)^2}{4}$.

\subsection{A Rosenzweig-MacArthur DDE model}
\label{sec:example_rosen}
\begin{figure}[b]
\centering
\subfigure[$\tau=0.01$.]{\includegraphics[width=0.32\textwidth]{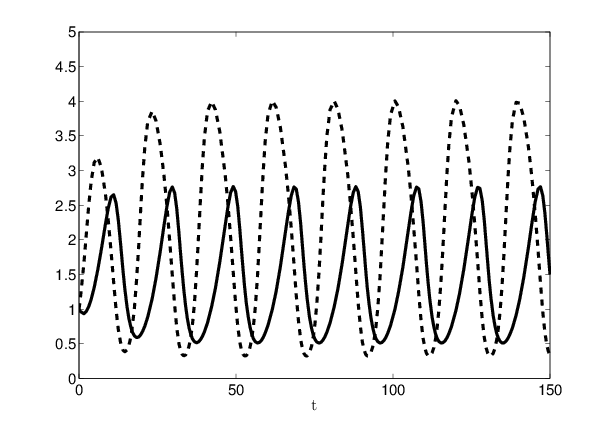}}
\subfigure[$\tau=0.3$.]{\includegraphics[width=0.32\textwidth]{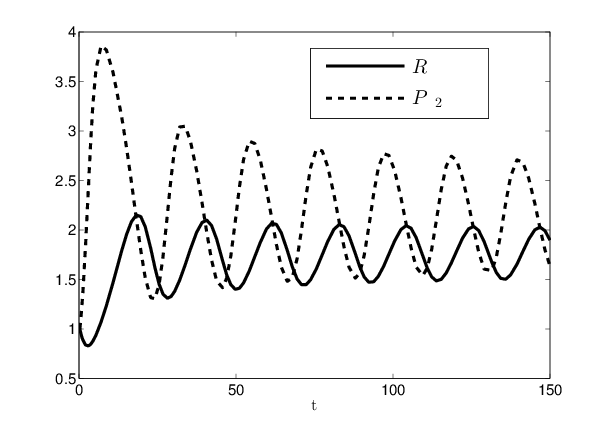}}
\subfigure[$\tau=0.4$.]{\includegraphics[width=0.32\textwidth]{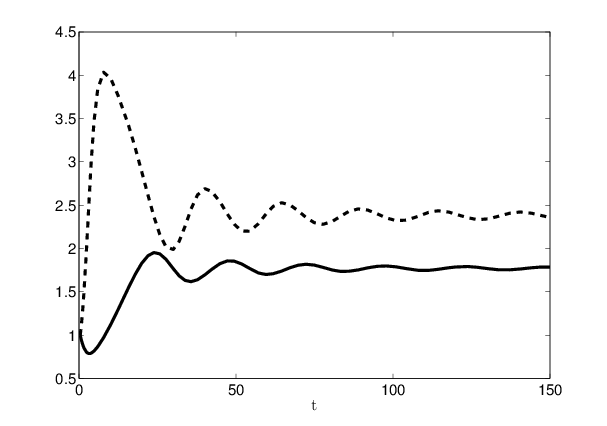}}
\caption{Effect of the delay on the solution of the system
\eqref{rosen_model_nondim}. Numerical
simulations were performed with the MATLAB\textregistered\, solver
\texttt{dde23}. For this simulations we choose $K=5,\,b=1,\mu=0.6,
\mu_1=0.4$ and three values of the delay $\tau$. We may observe stability
switches for $\tau \in [0,\tau_{max})$, with $\tau_{max}=0.3728$.}
\label{Fig:Rosen_StabSwitch}
\end{figure}

\begin{figure}[b]
\centering
 \subfigure{\includegraphics[width=0.47\textwidth]{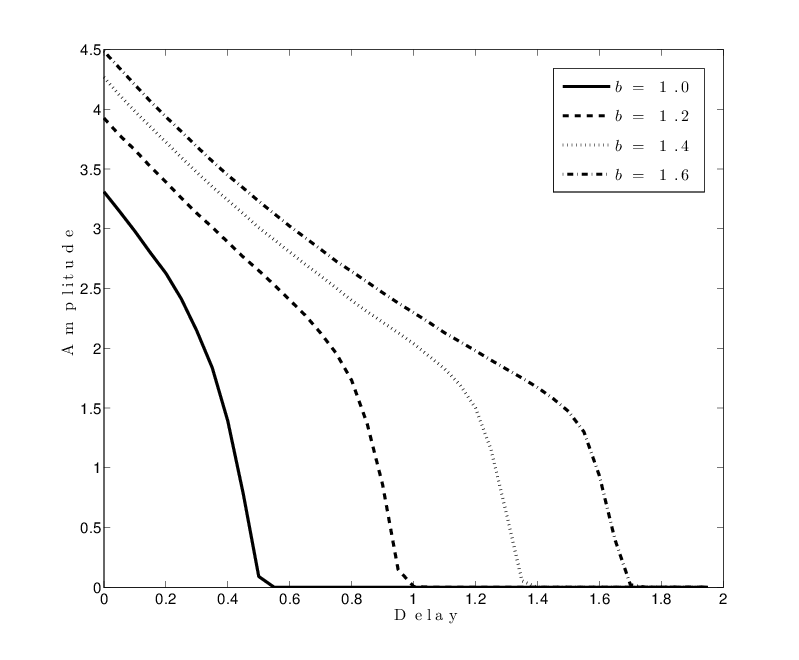}}
 \subfigure{\includegraphics[width=0.47\textwidth]{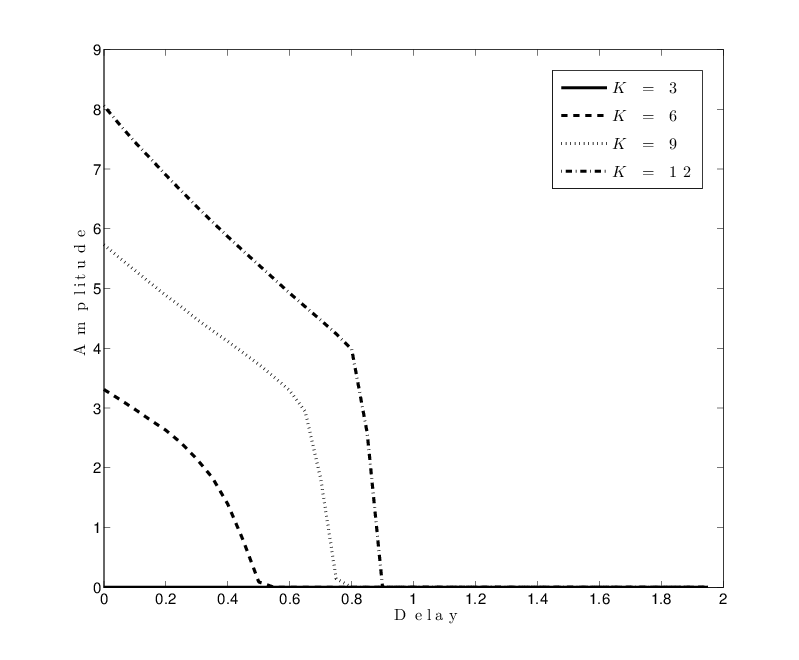}}\\
 \subfigure{\includegraphics[width=0.47\textwidth]{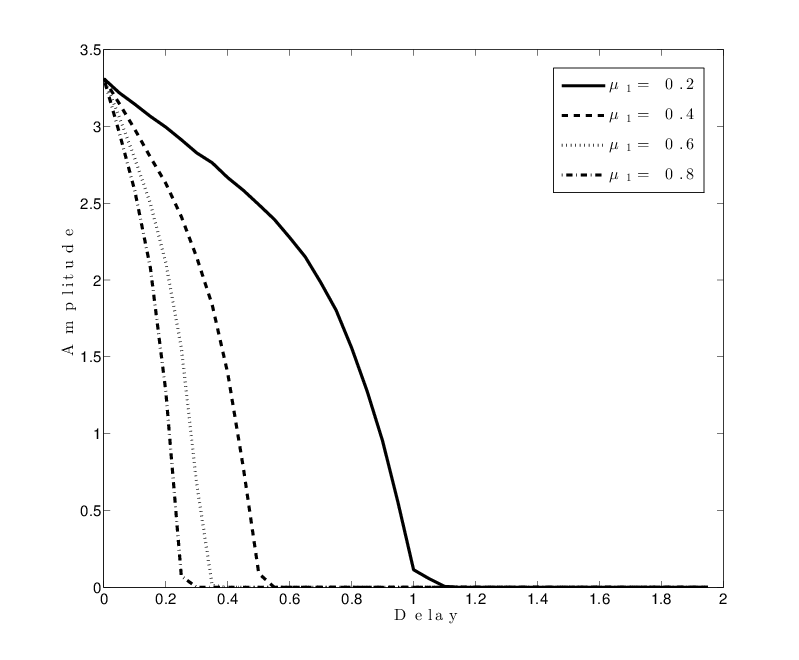}}
 \subfigure{\includegraphics[width=0.47\textwidth]{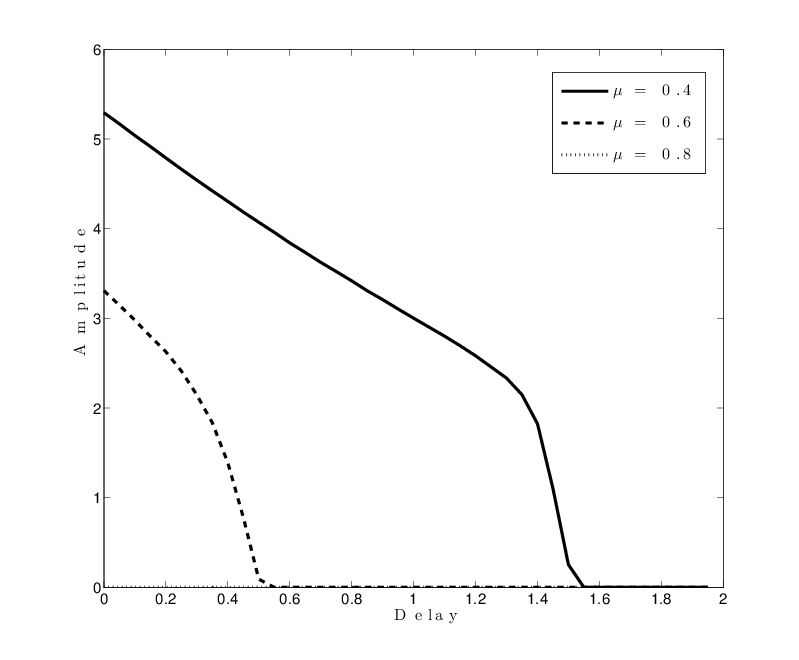}}
 \caption{Oscillation amplitudes of the $P_2$ population with respect to the
delay and one of the parameters. Unless otherwise indicated in the plot,
parameter values are $K=6,\,b=1,\mu=0.6, \mu_1=0.4$.}
 \label{fig:ampp}
 \end{figure}

An example for the class of problems \eqref{model1b_example} is given by a
delayed version of the
Rosenzweig-MacArthur model \eqref{sys:rosenzweig}. We choose the functions $h(t)
= h(R(t))=  R(t)/(1+R(t))$, $g(t) = g(R(t)) =  bh(t), \,b>0$, and
$\delta(P_1(t),P_2(t)) = \delta(P_2(t)) = P_2(t)$. Notice that these are all
$C^1$-functions and that $h(0)=0=g(0)$. As observed in the previous
section, the $P_1$-equation is not relevant for the dynamics of the whole
system. Hence, we focus on
the system
\begin{equation}
\begin{aligned}
R'(t) &= R(t) \left(1-\frac{R(t)}{K}\right) - b\frac{R(t)}{1+R(t)}P_2(t),\\
P_2'(t) &= -\mu P_2(t)  + b\frac{R(t-\tau)}{1+R(t-\tau)}P_2(t-\tau)e^{-\mu_1
\tau}.
\end{aligned}
\label{rosen_model_nondim}
\end{equation}
Assume that we have nonnegative continuous initial data \(R(\theta),
P_2(\theta)\) on \(-\tau \leq
\theta \leq 0\). Then it is easy to proof that the solutions \((R(t),P_2(t))\)
of \eqref{rosen_model_nondim} are nonnegative for all \(t\geq 0\)
\cite{Mohr2012}.\\
\ \\
In the case $\tau=0$, the dynamics of \eqref{rosen_model_nondim} is known
\cite{Rosenzweig1963}.
Beside
the trivial equilibrium, which always exists, system \eqref{rosen_model_nondim}
has a unique
coexistence
equilibrium \(E_C=(R^*,P_2^*)\) with
\begin{displaymath}
 R^* =   \frac{\mu}{b-\mu},\quad
P_2^* =  \frac{1}{b-\mu}\left(1- \frac{\mu}{K(b-\mu)}\right).
\end{displaymath}
The coexistence equilibrium is feasible only if $b>\mu$ and $K >
K^*_0=\frac{\mu}{b-\mu}$.
The point $E_C$ is locally asymptotically stable if \( K <
\frac{b+\mu}{b-\mu}\), a Hopf
bifurcation occurs at \(K^*_1 =\frac{b+\mu}{b-\mu}\) and the point becomes
unstable.\\
\ \\
In the case \(\tau >0\), for $b>\mu e^{\mu_1 \tau}$ and $K <
\frac{\mu}{be^{-\mu_1 \tau}-\mu}$, there is only one nontrivial equilibrium
$E_R=(K,0)$, which is stable for all
$\tau\geq 0$. For  $b>\mu e^{\mu_1 \tau}$ and $K > \frac{\mu}{be^{-\mu_1
\tau}-\mu}$
beside $E_R=(K,0)$, we find a unique coexistence point
$ E_C= \left(\frac{\mu}{b e^{-\mu_1 \tau}-\mu},
\frac{1}{b}\Bigl(1-\frac{R^*}{K}\Bigr)(1+R^*)\right)$, whose stability depends
on $\tau$. We determine the set $S_c$ in which stability switches can occur.
With \eqref{betagammaeps} we find
the values
\begin{displaymath}
\beta =\frac{2\mu}{K(b e^{-\mu_1 \tau}-\mu)} -e^{\mu_1 \tau} \left(\frac{\mu}{b}
+\frac{\mu}{Kb}\right),\qquad
\gamma = \mu e^{\mu_1 \tau},\qquad
\epsilon = \mu e^{\mu_1 \tau} \left[1-e^{\mu_1 \tau} \left(\frac{\mu}{b} +
\frac{\mu}{Kb}\right)\right],
\end{displaymath}
and we have that $ M(\tau) = \beta^2 \geq 0$. Hence, stability switches occur
for $\tau\geq0$ such that $N(\tau)>0$, or equivalently for
$\tau>0$ such that
\begin{displaymath}
\epsilon - 2\mu e^{\mu_1 \tau}\beta > 0 \quad \Leftrightarrow \quad
1+e^{\mu_1 \tau} \Bigl(\frac{\mu}{b} + \frac{\mu}{Kb}\Bigr) -\frac{4\mu}{K(b
e^{-\mu_1 \tau}-\mu)} > 0.
\end{displaymath}
That is, we want to find values $x>0$ such that
\begin{align}
x^2 + \frac{3b}{\mu (1+K)}x - \frac{Kb^2}{\mu^2 (1+K)} < 0 \label{R12}
\end{align}
holds. Zeros of (\ref{R12}) are
\begin{align*}
x_\pm = -\frac{3b}{2\mu (1+K)} \pm \sqrt{\frac{9b^2}{4\mu^2 (1+K)^2}
+\frac{Kb^2}{\mu^2 (1+K)}}.
\end{align*}
It is easy to see that \(x_-<0\). Thus $N(\tau)>0$ holds for all \(\tau \in
S_c=[0,\tau_{max})\), with $\tau_{max}=\ln(x_+)/\mu_1$. For $\tau \in
[0,\tau_{max})$
we can observe stability switches, for $\tau>\tau_{max}$ the stability of the
coexistence point
does not change when $\tau$ changes. An example is shown in Figure
\ref{Fig:Rosen_StabSwitch}. We observe that both $K$ and $\tau$ can be chosen as
bifurcation parameters for  \eqref{rosen_model_nondim}. For $\tau=0$ and $K\in
(0,K^*_0)$ the point
$E_R$ is stable, whereas for $K\in (K^*_0, K^*_1)$ the coexistence equilibrium
$E_C$ is locally
asymptotically stable and when the carrying capacity $K$
becomes larger than $K^*_1$ a limit cycle
appears.
The inclusion into the model of a maturation delay seems to shift forward the
destabilization of the
equilibria. That is, for
$\tau>0$ and $K\in (0,K^*_R(\tau))$, with $K^*_R(\tau)\leq K^*_R(0)=K^*_0$ the
point
$E_R$ is stable, whereas for $K\in (K^*_R(\tau), K^*_C(\tau))$ with
$K^*_C(\tau)\leq
K^*_C(0)=K^*_1$ the coexistence equilibrium $E_C$ is locally
asymptotically stable. However, for $K>K^*_C(\tau)$ an increase in the delay can
lead to damped
oscillations and again to a stable coexistence equilibrium, cf. Figure
\ref{Fig:Rosen_StabSwitch}. A last numerical test shows how periodic
solutions
depend on the delay and on the other parameters. We consider periodic solutions
and compute
oscillation amplitudes (over one period of the periodic solution). Figure
\ref{fig:ampp} shows
oscillation amplitudes for the solution $P_2$ in dependence of the delay and of
one other parameters. Observe that there is no periodic solution for very
large values of the delay. From a biological point of view, a very large
age-at-maturity implies that predators stay long in the
juvenile phase and it takes a long time for them to become active predators.
During this time the prey can increase and reach a certain density
such that the predation effects are not very relevant. This leads to a stable
coexistence of predators and prey, rather than the well-known oscillatory
dynamics.

